\documentclass[aps,nofootinbib,twocolumn,prl,pacs,class-pre]{revtex4} 
\usepackage{amsfonts}
\usepackage{amssymb}
\usepackage{graphicx}
\usepackage{pstricks}
\usepackage{epsfig}
\newcommand{\be}{\begin{equation}}
\newcommand{\ee}{\end{equation}}
\newcommand{\beq}{\begin{eqnarray}}
\newcommand{\eeq}{\end{eqnarray}}
\newcommand{\lapproxeq}{\lower .7ex\hbox{$\;\stackrel{\textstyle
<}{\sim}\;$}}
\newcommand{\gapproxeq}{\lower .7ex\hbox{$\;\stackrel{\textstyle
>}{\sim}\;$}}

\newcommand{\zl}{z_{\rm L}}
\newcommand{\zs}{z_{\rm S}}
\newcommand{\dls}{D_{\rm LS}}

\newcommand{\ds}{D_{\rm S}}
\newcommand{\simlt}{\mathrel{\hbox to 0pt{\lower
      3.5pt\hbox{$\mathchar"218$}\hss} \raise
      1.5pt\hbox{$\mathchar"13C$}}}

\newcommand{\simgt}{\mathrel{\hbox to 0pt{\lower
      3.5pt\hbox{$\mathchar"218$}\hss} \raise
      1.5pt\hbox{$\mathchar"13E$}}}

\begin{document}
\title{The necessity of dark matter in MOND within galactic scales}
\author{Ignacio Ferreras}\email{ferreras@star.ucl.ac.uk}
\author{Mairi Sakellariadou} \email{Mairi.Sakellariadou@kcl.ac.uk} 
\author{Muhammad Furqaan Yusaf} \email{Muhammad.Yusaf@kcl.ac.uk}
\affiliation{King's College London,
Department of Physics, Strand WC2R 2LS, London, U.K.}

\begin{abstract}
To further test MOdified Newtonian Dynamics (MOND) on galactic scales
-- originally proposed to explain the rotation curves of disk galaxies
without dark matter -- we study a sample of six strong gravitational
lensing early-type galaxies from the CASTLES database. To determine
whether dark matter is present in these galaxies, we compare the total
mass (from lensing) with the stellar mass content (from a comparison
of photometry and stellar population synthesis).  We find that strong
gravitational lensing on galactic scales requires a significant amount
of dark matter, even within MOND.  On such scales a 2~eV neutrino
cannot explain this excess matter -- in contrast with recent claims to
explain the lensing data of the bullet cluster. The presence of dark
matter is detected in regions with a higher acceleration than the
characteristic MONDian scale of $\sim 10^{-10}$m/s$^2$. This is a serious
challenge to MOND unless the proper treatment of lensing is
qualitatively different (possibly to be developed within a consistent
theory such as TeVeS).
\vspace{.2cm}

\noindent
PACS numbers: 
\end{abstract}

\maketitle


The standard ($\Lambda$CDM) cosmological paradigm is based on Cold
Dark Matter (CDM), a cosmological constant $\Lambda$, and classical
general relativity. Despite its enormous success and simplicity,
competing models have been proposed, the main reason
being the still unknown dark energy component and the undetectability
of dark matter. To explain the observed flat rotation curves,
Milgrom~\cite{milgrom} proposed MOdified Newtonian Dynamics (MOND),
based on the relation $f(|{\vec a}|/a_0){\vec a}=-{\vec
\nabla}\Phi_{\rm N},$ between the acceleration ${\vec a}$ and the
Newtonian gravitational field $\Phi_{\rm N}$.  The constant
$a_0\approx 10^{-10} {\rm m}/ {\rm s}^2$ is motivated by the
acceleration found in the outer regions of the galaxy where the
rotation curve is flat.  When $f$, assumed to be a positive smooth
monotonic function, equals unity,  usual Newtonian
dynamics holds, while when it approximately equals its argument, 
the deep MONDian regime applies.

MOND has been successful in explaining the dynamics of disk galaxies;
it is less successful for clusters of galaxies. It was
promoted~\cite{bek} to a relativistic field theory by introducing a
TEnsor, a VEctor and a Scalar field (TeVeS). TeVeS has been criticised
as lacking a fundamental theoretical motivation. Recently, it has been
argued~\cite{nickmairi} that such a theory can emerge naturally within
string models.

Here we calculate within MOND the deflection angles for two generic
density profiles and compare them with those predicted in standard
lensing. We calculate the mass of the lenses and estimate the amount of
dark matter required.  We find that despite the alternative
gravitational fall-off, the masses predicted by MOND are very similar
to those predicted within standard gravitational lensing theory. We
conclude that MOND within galactic scales needs a considerable amount
of dark matter.

\vspace{.2cm}
We consider a homogeneous and isotropic three-metric with the
density parameters ``tweaked'' to the values in a MONDian
cosmology.  The outcome of our lensing analysis depends only weakly on
the cosmology, for a reasonable range of cosmological
parameters. A different background cosmology mainly results in 
the change of the critical surface mass density\cite{cosmo}.

Assuming that the deflection of photons is twice that of non-relativistic
particles and that the photon path is nearly linear, the deflection
angle $\alpha$ as a function of the impact parameter $b$ can be written, 
for a given cumulative mass profile $M(<r)$, as (see e.g.~\cite{mt01}):
\beq
 \alpha(b)=-\frac{4Gb}{c^2}\int^\infty_0 f^{-1/2}
\left(\frac{GM(<\sqrt{b^2+z^2})}{[b^2+z^2]a_0}\right)\nonumber\\
\times \frac{M(<\sqrt{b^2+z^2})}{[b^2+z^2]^{3/2}}\ {\rm d}z~. 
\label{eq:deflang} 
\eeq 
When the function $f(x)$ in the integrand is removed, we recover the
expression of the deflection angle in standard lensing. The function
$f(x)$ ``modulates'' this deflection along the path of the particle
depending on the ratio between the local acceleration, 
 $GM(<r)/r^2$, and the MONDian characteristic acceleration, $a_0$.
We will henceforth use Eq.~(\ref{eq:deflang}) to calculate the
deflection angle. In standard lensing $f(x)$ is set to unity, while in
MOND we first adopt~\cite{5} $f(x)=x[1+x^2]^{-1/2}$.

\begin{figure*}
\begin{minipage}{14cm}
  \includegraphics[width=0.4\linewidth]{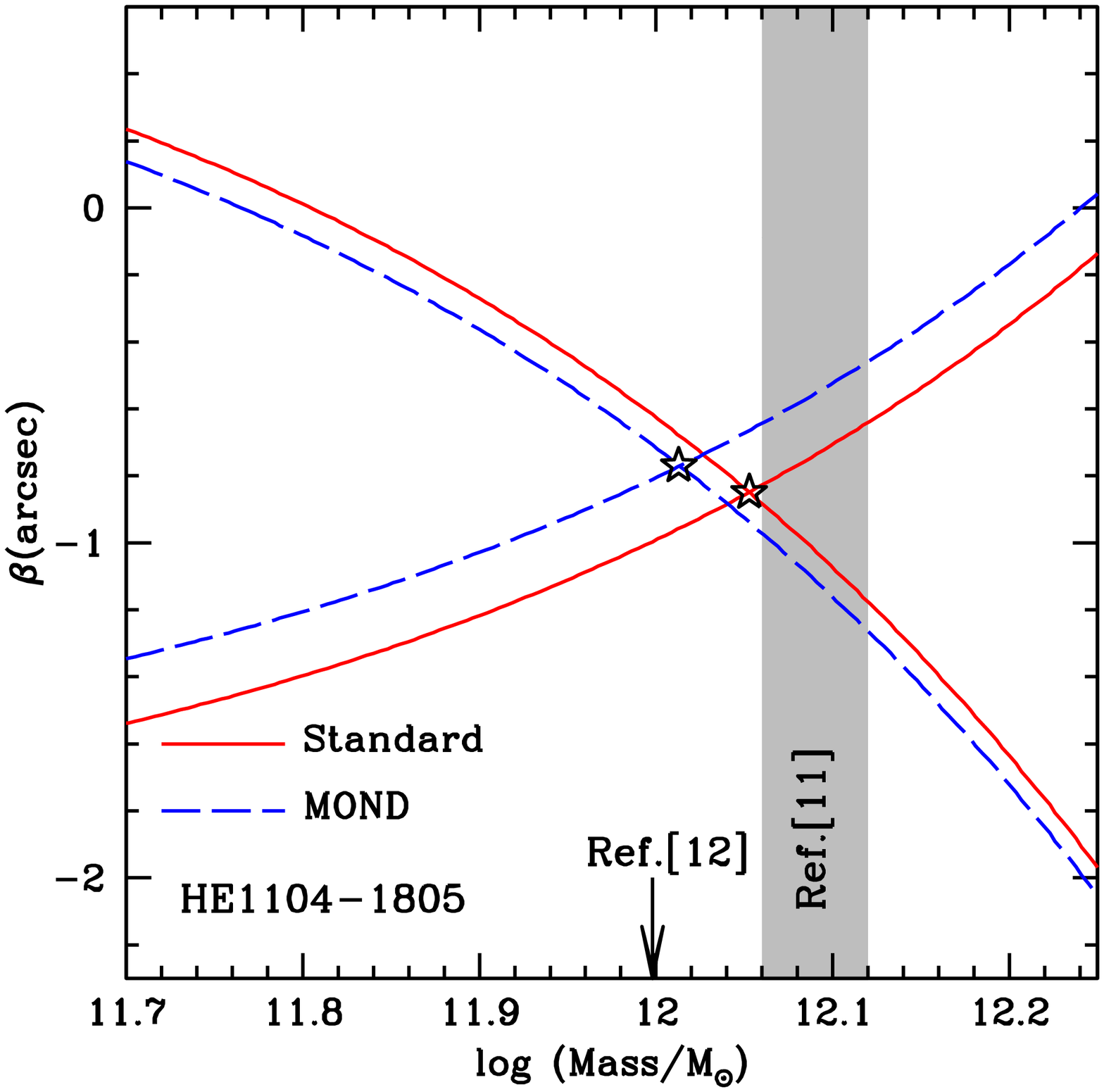}
  \includegraphics[width=0.4\linewidth]{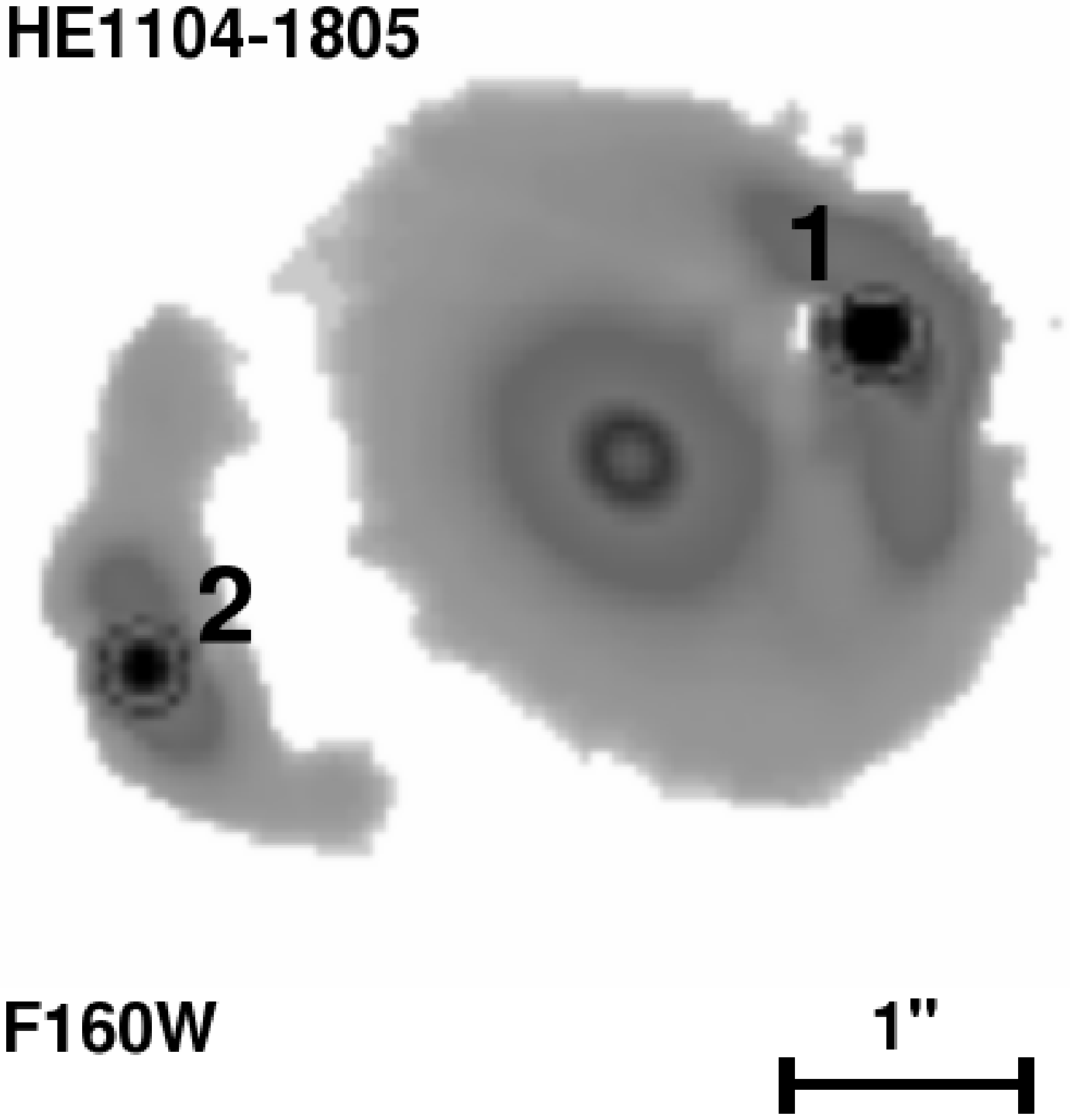}
\caption{{\sl Left}: Graphical representation of the lens equation in
       standard lensing (solid lines) and MOND (dashed lines).
       Each line corresponds to one of the two images of
         the background source. The distant one (number 2)
       corresponds to the lower set of lines (i.e. a more discrepant
       result between standard and MONDian lensing). The intersection
       point of the two lines gives the position of the source and the
           total mass (Hernquist profile assumed). {\sl Right}:
       NIR HST/NICMOS grey-scale image of the lensing system (from the
       CASTLES database).} 
  \label{fig:HE1104}
\end{minipage}
\end{figure*}

We compare observations of strong lensing systems (which are most
often elliptical galaxies) with realistic mass profiles. Spherical
symmetry is assumed. In addition to the ``no-dark-matter''
interpretation of the rotation curves in disk galaxies, we assume that
in MOND the stellar mass content represents the full mass budget;
the contribution of other baryonic components such as gas or dust is
minimal in early-type systems.  Their characteristic surface
brightness profile can be represented by a Hernquist 3-D density
profile~\cite{hern90}. The cumulative mass profile is $M(<r)=M
r^2/(r+r_{\rm h})^2$, where $M$ is the total mass of the galaxy and
$r_{\rm h}$ the core length scale, related to the projected 2-D
half-mass radius by $R_{\rm e}=1.8153\ r_{\rm h}$.  This density model
has a logarithmic slope $({\rm d}\log\rho)/({\rm d}\log r)\propto -1$
towards the centre, changing to $-4$, as $r\rightarrow\infty$. This is
our first model.

The Navarro-Frenk-White (NFW) profile~\cite{NFW} is our second
model. The cumulative mass profile diverging logarithmically, we
assume a truncation radius $r_{\rm virial}$.  This profile has two
free parameters, the core length scale $r_{\rm s}$, and the virial
radius. Their ratio is the concentration ${\cal C}$.  Cosmological
simulations~\cite{wechs06} suggest concentrations on galaxy scales 
to be ${\cal C}\sim 10$. Denoting by $x$ the ratio
 $x\equiv r/r_{\rm virial}$, the
cumulative mass function of the NFW profile reads 
\be 
M(<r)=M\frac{\ln (1+{\cal C}x)-\frac{{\cal C}x}{1+{\cal C}x}}{\ln
(1+{\cal C})-\frac{{\cal C}}{1+{\cal C}}}~.  
\ee

The lens equation $\beta = \theta - \alpha (\theta)\dls/\ds$ relates
the actual position of the background source $\beta$, with the
position $\theta$ of the images. For a given cosmological model, the
angular diameter distances from the lens to the source, and from the
observer to the source, $\dls$ and $\ds$ respectively, are
obtained from the observed redshifts. The deflection angle $\alpha$
depends on the mass profile of the system and the impact parameter.  A
characteristic aspect of strong gravitational lensing is that one
image appears inside the Einstein radius $r_{\rm E}$ and the other one
outside. The difference between MONDian and standard lensing lies mostly 
in the position of the image outside $r_{\rm E}$.

\begin{table*}[h]
\begin{minipage}{15cm}
\caption{Mass estimates (in $10^{10}M_\odot$ units) for $\Lambda$CDM
cosmology: $(\Omega_{\rm m},\Omega_\Lambda,\Omega_{\rm
k})=(0.3,0.7,0)$. The masses in brackets correspond to the open
cosmology of Ref.~\cite{zhao06}: $(\Omega_{\rm
m},\Omega_\Lambda,\Omega_{\rm k})=(0.03,0.36,0.51)$. 
}
\begin{tabular}{l||rr|rr|cc|c}
\hline &\multicolumn{2}{c}{Hernquist} & \multicolumn{2}{c}{NFW (${\cal
 C}$=10)} & \multicolumn{2}{c}{Ref.~\cite{fsw05} (standard)} &
 Ref.~\cite{zhao06}\\ \textbf{Lens} & {\hfill standard\hfill} &
 {\hfill MOND\hfill} & {\hfill standard\hfill} & {\hfill MOND\hfill} &
 {\hfill standard\hfill} & {\hfill M$_{\rm STAR}$\hfill} & {\hfill
 MOND\hfill}\\ \hline Q0142-100 & $ 32.37( 34.58)$ & $ 29.28( 31.56)$
 & $ 29.67( 31.70)$ & $26.63( 28.74)$ & $24.9^{31.7}_{20.2}$ &
 $20.9^{30.8}_{13.0}$ & $29.9$\\ HS0818+1227 & $ 50.99( 54.03)$ & $
 46.31( 49.50)$ & $ 48.14( 51.01)$ & $43.38( 46.42)$ &
 $67.4^{73.6}_{60.7}$ & $16.2^{21.2}_{12.6}$ & --\\ FBQ0951+2635 & $
 4.07( 4.16)$ & $ 3.82( 3.91)$ & $ 3.28( 3.35)$ & $ 3.07( 3.14)$ &
 $4.7^{5.7}_{3.6}$ & $1.1^{2.1}_{0.5}$ & $3.6$\\ BRI0952-0115 & $
 7.33( 5.25)$ & $ 6.62( 4.80)$ & $ 8.37( 3.42)$ & $ 7.48( 3.10)$ &
 $4.5^{4.9}_{4.2}$ & $3.5^{4.0}_{2.7}$ & $4.3$\\ Q1017-207 & $ 9.93(
 10.95)$ & $ 9.04( 10.04)$ & $ 9.57( 10.55)$ & $ 8.64( 9.61)$ &
 $4.8^{6.2}_{4.5}$ & $4.3^{13.0}_{1.4}$ & $14.7$\\ HE1104-1805 &
 $112.93(123.11)$ & $103.17(113.25)$ & $ 89.63( 97.71)$ & $81.28(
 89.29)$ & $122.0^{130.0}_{115.0}$ & $22.8^{51.2}_{12.7}$ & $99.6$\\
 \hline
\end{tabular}

\label{tab:mass}
\end{minipage}
\end{table*}

Figure~\ref{fig:HE1104} illustrates our methodology in estimating the
masses of galaxies from lensing data. HE1104-1805 is extracted from
the CfA-Arizona Space Telescope Survey (CASTLES~\cite{rus03})
sample. It consists of a galaxy at redshift $\zl=0.73$ with a
background QSO at $\zs=2.32$. A grey-scale map of the HST/NICMOS F160W
image is shown on the right panel, retrieved from the CASTLES
web-page~\footnote{ http://cfa-www.harvard.edu/castles/.}.
 This is a double system with the image
positions located at 2.09 and 1.10~arcsec on either side of the
lensing galaxy. The left panel of Fig.~\ref{fig:HE1104} shows the
correlation between the actual position $\beta$ of the QSO, and the
total mass of the lensing galaxy, assuming a Hernquist profile with
the projected 2-D half-mass radius being equal to the observed
half-light radius of the lensing galaxy. Each set of lines -- dashed
(MOND) or solid (standard lensing theory) -- are the results for each
image. The compatible solution corresponds to the crossing of the
lines, shown in the figure with a star symbol. This gives the true
position of the source and the mass of the galaxy. For comparison, the
values from Refs.~\cite{fsw05} (for conventional lensing theory) and
~\cite{zhao06} (for MOND) are given as a shaded region and an arrow,
respectively.

\begin{figure*}
\begin{minipage}{14cm}
\includegraphics[width=0.8\linewidth]{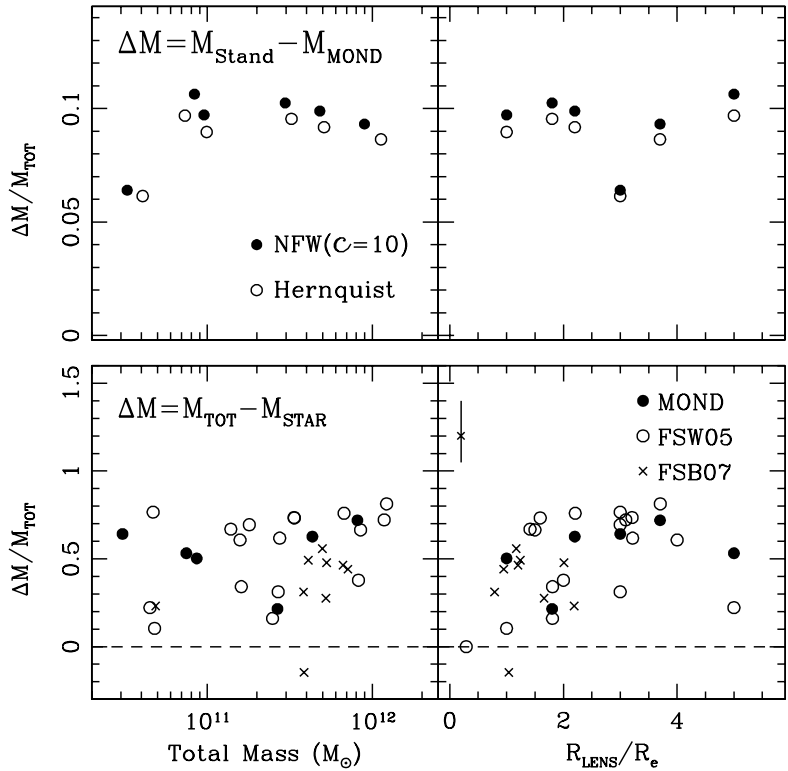}
\caption{{\sl Top}: Difference between conventional and MOND masses
for a NFW model with ${\cal C}=10$ (filled dots) and a Hernquist
profile (hollow dots).  The ratio $\Delta M\equiv M_{\rm std}-M_{\rm
MOND}$ is shown as a function of total (standard) mass (left panel)
and the ratio between the average lens separation over which lensing
masses can be reliably measured, and the observed half-light radius
(right panel).  {\sl Bottom}: Contribution of dark matter to the
total mass budget from a comparison between MONDian lensing and
stellar mass. for a NFW model with ${\cal C}=10$ (filled dots). We
also include more detailed non-parametric conventional mass estimates
of strong lenses from Refs.~\cite{fsw05} and \cite{fswb07}.}
\label{fig:diff}
\end{minipage}
\end{figure*}

Table~I compares our mass estimates with the MONDian analysis of
Ref.~\cite{zhao06} and with the standard non-parametric approach of
Ref.~\cite{fsw05} (where spherical symmetry is not assumed).  The
masses are quoted in units of $10^{10} M_\odot$ for a $\Lambda$CDM
cosmology and, in brackets, for the open cosmological model of
Ref.~\cite{zhao06}.  A Chabrier~\cite{chab03} initial mass function is
considered for the stellar masses quoted from Ref.~\cite{fsw05}. The
resulting synthetic population, constrained by the photometry of the
lensing galaxy in the optical (F814W) and NIR (F160W) passbands, is
used to determine the stellar mass content. The sample studied here
comprises only double systems to be suitable for a 1-D approximation
of the lens and serves to show the differences between MOND and
standard lensing.

Table~I shows a small difference in the mass estimates between the
two different cosmologies considered here, despite their density
parameters being quite different. This is because the angular distance
is mostly unaffected by the change in the parameters.  There are also
some noticeable differences between a Hernquist and a NFW (${\cal
C}=10$) model for the distribution of mass in the lensing
galaxy. Nevertheless, the differences found are not large enough to
affect our conclusions. One could always argue for a Hernquist profile
as this is the model that a baryon-only MONDian cosmology would
favor, given that the projected mass distribution resembles the
typical de~Vaucouleur profile of early-type galaxies. However, recent
lensing work on clusters, most noticeably the bullet
cluster~\cite{bullet} has been used to postulate a 2~eV neutrino which
would be important on scales of galaxy clusters, not on galactic
scales~\cite{sand07}.  We present the NFW profile, to illustrate the
robustness of our claims in rejecting the hypothesis of a 2~eV
neutrino.

The top panels of Fig.~\ref{fig:diff} compares the mass estimates
between standard theory and MOND for both density profiles: Hernquist
(hollow dots) and NFW with ${\cal C}=10$ (filled dots). The mass
differences are shown as a function of conventionally calculated mass
(left panel) and R$_{\rm LENS}$/R$_e$ (right panel). The difference
between the conventional theory and MONDian predictions stays mostly
within 10\%. This is especially noteworthy in systems with R$_{\rm
LENS}/$R$_e\simgt 2$. Notice that the lensed images probe
accelerations slightly above the MONDian threshold. For instance, in
lens HE1104-1805 (figure~1), image 2 (right panel) is located on the
lens plane at a point with a local acceleration of 
$4.5\times 10^{-10}$m/s$^2$ (using the MOND mass estimate in table~1 
for a Hernquist profile), which explains why the difference between the
solid (standard lensing) and the dashed lines (MOND) in the leftmost
panel is so small.

The bottom panels of Fig.~\ref{fig:diff} puts this result in context
with the need for dark matter. The figure compares MONDian lensing
mass with stellar mass (solid dots). Our 1-D estimates are compared
with more detailed non-parametric models from Refs.~\cite{fsw05} and
\cite{fswb07}. A typical error bar from these estimates is also
shown. Even though some of the systems can be compatible with no dark
matter, the MONDian analysis presented here finds in most cases the
need for dark matter at a level around $M_{\rm DM}/M_{\rm STAR}\sim$
0.5--2. Given that the dust and gas content in early-type galaxies
corresponds to a fraction of the stellar mass, we infer the need for
dark matter even within MOND. Our analysis shows that 
dark matter in early-type systems appears in regions with different
absolute accelerations compared to disk galaxies. Hence, a theory
with a fixed absolute acceleration (such as MOND) cannot explain both
early- and late-type systems.

The form of the function $f(x)$, which varies smoothly from the deep
MONDian to the standard regime is an extra source of uncertainty in
the MONDian mass estimates. If $f(x)$ varies too slowly, lingering
close to the conventional regime for too long, MONDian mass
predictions are too high, while if $f(x)$ falls quicker to the MONDian
limit, the need for dark matter would diminish.  There is no precise
way to determine the exact form of this function. From galactic
rotation curves some restrictions can be placed on its form, but there
still exists a degree of freedom.  Varying the form of $f(x)$, it was
found~\cite{zhao06} that the predicted masses are not affected
considerably and that many of the lenses still give a high dark matter
content.  Here, we considered two alternatives for the acceleration
function, namely $f(x)=x/(1+x)$ and $f(x)=1-e^{-x}$. The MOND mass
estimates are lowered by less than 10\%. Note that one could
manufacture a function $f(x)$ such that MOND can be successful without
dark matter, however such artificially made functions would 
disregard the data from rotation curves.

Another possible source of uncertainty lies in the absolute value of
the acceleration scale $a_0$. One can increase $a_0$ by a factor $2$
and still be compatible with the rotation curve data~\cite{sm02}. In
our case, the mass estimates are lowered by about 10\%. A
combination of a higher $a_0$ and a shallower function $f(x)$ can
result in mass estimates lower than our fiducial MOND estimates by
about 25\% which would still not be large enough to make dark matter
unnecessary.

\vspace{.2cm} 
In this paper we have compared mass estimates for a set
of early-type lensing galaxies using both standard lensing analysis
and MOND.  We used two density profiles, the NFW profile and the
Hernquist profile.  We used the lensing equations to predict the mass
of a system from the image positions for a 1-D model (spherical
symmetry).  Besides the standard paradigm $\Lambda$CDM cosmology,
other recent alternatives from the literature were considered,
including the possible solution presented in Ref.~\cite{sko06} where
the addition of massive neutrinos allows a cosmology of $(\Omega_{\rm
m},\Omega_\Lambda,\Omega_{\rm k})=(0.22,0.78,0)$ to give an acceptable
fit to both the CMB angular power spectrum as well as the
high-redshift supernova data. For our purposes, any of the cosmologies
discussed give very similar mass estimates, a result which should not
come as a surprise since the observational constraints mostly impose
limits on the luminosity and angular diameter scales.

We tested MOND by looking at a set of strong gravitational
lensing early-type galaxies from the CASTLES survey. The masses
predicted in the framework of conventional theory are very close to
those from MONDian lensing, even for galaxies observed out to a few
effective radii. Comparing the stellar mass content from a comparison
of the observed optical and NIR photometry with stellar population
synthesis models we found that a very similar amount of dark matter is
needed in both conventional and MOND analysis. This result is in
remarkable contrast with the recent attempts to explain the lensing
data on cluster scales by introducing a 2~eV
neutrino~\cite{sand07}. This component can cluster on Mpc scales but
should not cluster on galactic scales to keep the analysis of the
rotation curves of disk galaxies unchanged. However, our lenses, which
do require dark matter, are studied over length scales comparable to
those of the rotation curve analysis. We therefore conclude that
either lensing must work in a qualitatively different way within MOND
(or more correctly the covariant ``parent'' theory, such as TeVeS) or
dark matter should be considered within MOND even on galactic
scales.

\vspace{.2cm} \acknowledgments It is a pleasure to thank Prasenjit
Saha and HongSheng Zhao for discussions.  I.\ F. was supported in part
by the Nuffield Foundation.  M.\ S. was supported in part by the
European Union through the Marie Curie Research and Training Network
UniverseNet (MRTN-CN-2006-035863).

\end{document}